# Near-infrared optical properties and proposed phase-change usefulness of transition metal disulfides

Akshay Singh,[1] Yifei Li,[1] Balint Fodor,[2] Laszlo Makai,[2] Jian Zhou,[3] Haowei Xu,[4] Austin Akey,[5] Ju Li,[4] R. Jaramillo[1*]

1. Department of Materials Science and Engineering, Massachusetts Institute of Technology, Cambridge, MA, 02139, USA.

2. Semilab Semiconductor Physics Laboratory Co. Ltd., Budapest, Hungary

3. Center for Advancing Materials Performance from the Nanoscale, State Key Laboratory for Mechanical Behavior of Materials, Xi'an Jiaotong University, Xi'an 710049, China

4. Department of Nuclear Science and Engineering, Massachusetts Institute of Technology, Cambridge, MA, 02139, USA.

5. Center for Nanoscale Systems, Harvard University, Cambridge, Massachusetts 02138, USA

* rjaramil@mit.edu

*Abstract*

The development of photonic integrated circuits would benefit from a wider selection of materials that can strongly-control near-infrared (NIR) light. Transition metal dichalcogenides (TMDs) have been explored extensively for visible spectrum opto-electronics, but the NIR properties of these layered materials have been less-studied. The measurement of optical constants is the foremost step to qualify TMDs for use in NIR photonics. Here we measure the complex optical constants for select sulfide TMDs (bulk crystals of $MoS_2$, $TiS_2$ and $ZrS_2$) via spectroscopic ellipsometry in the visible-to-NIR range. Through Mueller matrix measurements and generalized ellipsometry, we explicitly measure the direction of the ordinary optical axis. We support our measurements with density functional theory (DFT) calculations, which agree with our measurements and predict giant birefringence. We further propose that TMDs could find use as photonic phase-change materials, by designing alloys that are thermodynamically adjacent to phase boundaries between competing crystal structures, to realize martensitic (*i.e.* displacive, order-order) switching.

*1. Introduction: TMDs for NIR integrated photonics*

Integrated photonics offers a way to create all-optical circuits to reduce the power needed to move and process massive data flows, and to move beyond von Neumann computing by incorporating paradigms such as compute-in-memory and deep learning [1,2]. Photonic integrated circuits are usually made on silicon, which has low-loss in the near-infrared (NIR) [3]. Essential for photonic circuits are active materials that can modulate the phase and amplitude of light to perform switching, logic, and signal processing; if these changes are non-volatile, then they can also be used for memory. The most well-established active materials for photonics do not interact strongly with NIR light, and therefore require a large interaction volume and are not suitable for aggressively-miniaturized integrated photonic circuits [4]. The interaction length required to produce a substantial modulation of the optical phase is $L \sim \lambda_0/\Delta n$, where $\lambda_0$ is the free-space wavelength, and $\Delta n$ is the refractive index change. For instance, LiNbO$_3$ produces $\Delta n$ of $\mathcal{O}(0.001)$ at typical supply voltages for electro-optic modulation, and therefore requires an interaction length $L > 1$ mm. A leading class of candidate active materials for integrated photonics that can strongly modulate NIR light on a sub-µm length scale are so-called phase-change chalcogenides, such as those found in the Ge-Sb-Te (GST) system [5]. These materials operate by switching between crystalline and amorphous phases via time-temperature processing, which can be effected with light or electrical stimuli on a timescale of tens of nanoseconds, and which produce refractive index changes $\Delta n > 1$ [6]. Unfortunately these materials suffer from high optical losses in the NIR, and the melt-quench and recrystallization processes require large energy inputs and are limited to the sub-GHz operation [7,8]. There is a need to expand the selection of materials available for phase-change functionality in the NIR for integrated photonics.

Transition metal dichalcogenides (TMDs) are layered materials (van der Waals-bonded solids) with intriguing physical properties that include layer-number-dependent band gap, electron pseudospin, exciton and trion excitations, chemical tunability, catalytic action, polymorphism and phase-change behavior, and strong above-band gap light absorption [9–17]. The NIR and below-band gap optical properties of TMDs have been little-studied [18–21]. TMDs interact strongly with light, and are expected to feature low-loss for below-band gap wavelengths. Polymorphism suggests that transitions between structural phases, such the trigonal prismatic 2H and octahedral 1T (or distorted 1T' and 1T$_d$), may be useful for optical switching [22,23]. Transitions between

2H and 1T can be described by a simple translation of a plane of chalcogen atoms (a martensitic transformation) [24]. The layered, van der Waals crystal structure suggests that martensitic transformation strain may be low, which is beneficial for switching energy and fatigue. Phase-change functionality (*i.e.* transformation between layered polymorphs) at room-temperature has been demonstrated for TMDs including $MoTe_2$ and $(Mo,W)Te_2$ and may be expected to feature large $\Delta n$, although this has not been studied [23,25–27]. Here we focus on sulfide TMDs, because they have the largest band gap (relative to selenides and tellurides), and therefore offer the largest spectral range for low-loss, below-band gap operation. Unfortunately, the energetic cost of switching between phases is also highest for pure sulfides (relative to selenides and tellurides) [25,28]. We propose that alloying sulfide TMDs with different equilibrium structural phases, such as 2H $MoS_2$ and 1T $TiS_2$, could enable low-power switching. Specifically, these alloyed materials can be designed to be adjacent to a thermodynamic phase boundary.

The layered structure of TMDs leads to substantial anisotropy in physical properties, including refractive index (**Figure 1a**). TMDs have high in-plane symmetry and are expected to be birefringent, with the ordinary optical axis corresponding to the out-of-plane direction (**Figure 1b**); lower-symmetry phases such as 1T' may be trirefringent, although the in-plane anisotropy is likely dwarfed by the difference between the in-plane and out-of-plane indices. A large birefringence can be exploited for photonic components such as modulators, phase shifters and polarization converters [29]. Early work on bulk TMD crystals has suggested substantial anisotropy in reflectivity [30].

Here we measure the complex relative permittivity ($\epsilon = \epsilon_1 - i\epsilon_2$) of 2H-$MoS_2$, 1T-$ZrS_2$, and 1T-$TiS_2$ bulk crystals in the visible-to-NIR region (300 - 2100 nm), using spectroscopic ellipsometry (SE). We find that $\epsilon$ in the NIR cannot be simply extracted by extrapolating from visible light measurements, and requires explicit non-model-dependent measurements. We find that spectroscopic measurements must account for the presence of native oxide to avoid overestimating NIR loss. Our Mueller matrix measurements are consistent with symmetry expected of uniaxial, layered materials, with the optical axis out-of-plane. We support our measurements with density functional theory (DFT) calculations, which predict a giant birefringence. We suggest a new paradigm to use TMD alloys as active phase-change materials for integrated photonics, and use DFT to calculate alloy phase stability.

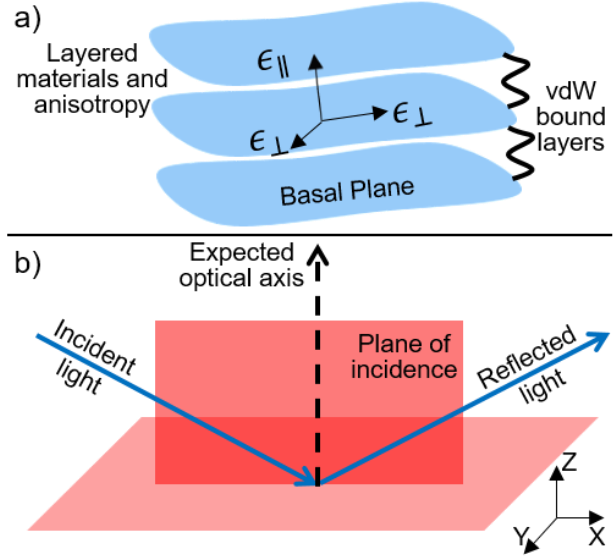

**Figure 1:** Layered structure and optical geometry of TMDs. a) The structure of layered TMDs results in optical birefringence. b) Schematic of plane of incidence defined by incident and reflected light, along with the coordinate system (XYZ). The expected position of optical axis for uniaxial TMDs is indicated (dotted lines).

## 2. Techniques to determine the NIR properties of TMDs: Ellipsometry and DFT

Ellipsometry is a non-destructive optical technique that is widely-used to measure the optical constants of thin films and bulk crystals [31,32]. The technique involves measuring the ellipsometric ratio ($\rho$) of the amplitude reflection Fresnel coefficients for P- and S-polarized light ($r_p$ and $r_s$, respectively) incident on a smooth surface. $\rho$ is written as

$$\rho = \frac{r_P}{r_S} = \tan(\psi) \exp(i\Delta), \qquad (1)$$

where $\psi$ and $\Delta$ are the ellipsometric angles. $\rho$ is a complex number, and can be used to directly calculate real and imaginary optical constants, without relying on Kramers-Kronig (KK) relations. Just relying on KK constrained data can result in erroneous results, due to extension of KK integral into spectral regions where measurements were not performed.

Equation (1) assumes no cross-polarization components, which can result from certain asymmetries and structure in the material. A more general form of the reflection matrix can be defined with help of Jones matrix ($S$):

$$S = \begin{bmatrix} r_{pp} & r_{ps} \\ r_{sp} & r_{ss} \end{bmatrix} \Rightarrow \rho = \begin{bmatrix} \rho_{pp} & \rho_{ps} \\ \rho_{sp} & 1 \end{bmatrix} \quad , \tag{2}$$

where the cross-polarization components are measured by $\rho_{ps}$ and $\rho_{sp}$. Experimentally, these components can be measured by using Mueller matrix (MM) formalism (a.k.a. generalized ellipsometry), which measures the complete dielectric polarization of a material [33,34]. The symmetry of the MM provides information about the optical symmetry of the material, such as the orientation of the birefringent optical axis relative to the incidence plane of the measurement [33,35]. Further, by using multiple angles of incidence (AOI), $\epsilon_\perp$ and $\epsilon_\parallel$ can be uncoupled for an anisotropic material [36]. Previous measurements have partially-characterized 2H-MoS$_2$ bulk crystals and thin films, with limited information in the NIR [37–41]. Much less information is available for 1T-ZrS$_2$ and 1T-TiS$_2$ [42].

We use thick TMD crystals to directly measure the dielectric properties in the NIR, without relying on models or extrapolation from visible-light measurements. For sufficiently-thick crystals the light is absorbed completely, and there are no reflections from the back surface that can mix polarizations and cause measurement errors. We obtain crystals from 2D Semiconductors (MoS$_2$, ZrS$_2$, and TiS$_2$ synthesized by chemical vapor transport) and the Smithsonian Institution (MoS$_2$, naturally-occurring molybdenite, catalog number NMNH B3306). We perform measurements on the mirror-like faces of the as-received crystals, without any surface processing steps; below we discuss how we account for the inevitable native oxide layer. We use two different ellipsometry instruments (Semilab SE-2000, and Woollam UV-NIR Vase) to ensure repeatability. Both instruments enable ultraviolet-to-NIR measurements with a single experimental setup (see Supplementary for more details), and allow the use of focusing optics. The size of available flat, mirror-like facets varies between samples (~ 1 mm for MoS$_2$ and ZrS$_2$, and ~5 mm for TiS$_2$), and necessitates use of focusing optics (~ 300 $\mu m$ spot size for 70° AOI) for input and output light. We compared data taken with and without focusing optics for TiS$_2$, and the results are similar (see Supplementary).

The procedure for extracting $\epsilon$ from measurements of $\rho$ is greatly simplified by using bulk crystal samples, as opposed to thin films. For a bulk, isotropic material, $\epsilon$ can be directly calculated as

$$\epsilon = sin^2(\Phi) * \left(1 + tan^2(\Phi) * \left(\frac{1-\rho}{1+\rho}\right)^2\right), \quad (3)$$

where $\epsilon = \epsilon_1 - i\epsilon_2$, and $\Phi$ is the AOI relative to the surface normal direction [31]. In the presence of reflections from the back of the sample or presence of oxide overlayers, Equation (3) cannot be used, and measurements of $\epsilon$ in the transparent spectral regions (*i.e.* below band gap) are more complicated and less accurate. $\epsilon$ calculated using Equation 3 is called the "effective" permittivity and corresponds to a model of a pristine material interface with air, without a native oxide or any other overlayer. The presence of a native oxide can produce substantial errors including an overestimation of optical loss, as discussed below and in the Supplementary.

We use the results of DFT electronic structure calculations to predict dielectric functions using the random phase approximation, as described in previous work and in the Supplementary [43,44]. We perform DFT calculations using the Vienna ab-initio simulation package (VASP), version 5.4 [45–48]. We treat the core and valence electrons by the projector-augmented plane-wave method, and we approximate the exchange-correlation interaction by the generalized gradient approximation functional, implemented in the Perdew-Burke-Ernzerhof form [49,50]. The energy minimization and force convergence criteria are $10^{-7}$ eV and $10^{-3}$ eV/Å, respectively.

## *3. Results: Complex optical constants and birefringence*

In **Figure 2** we present the experimentally-measured and calculated in-plane complex permittivity for MoS$_2$, TiS$_2$ and ZrS$_2$. Of particular relevance for NIR photonics, $\epsilon_1$ is large below the band gap of MoS$_2$ and ZrS$_2$ (indirect $E_g$ = 1.1 and 1.6 eV, respectively, indicated by solid gray lines); TiS$_2$ is a semimetal, with a band gap due to dimerization that is estimated to be in the range 0.3 - 0.5 eV at room temperature [20,42,51]. The DFT calculations match fairly well the experimental data, both in magnitude and in spectral position of individual features. The A, B and C excitons of MoS$_2$ are well-resolved [52]. For TiS$_2$, the experimentally-observed peaks match in energy but are substantially broader than those calculated by DFT, and are qualitatively similar to previously-reported measurements of TiSe$_2$ [19]. In ZrS$_2$, the strongest direct gap excitonic oscillators (2.5 eV and 2.9 eV) are observed in both experiment and theory, although the experimental data doesn't show the lowest-lying indirect gap transition near $E_g$. The 2.5 eV and 2.9 eV transitions are sufficiently close in energy to create zero-crossing of $\epsilon_1$ near 3.2 eV, which is not seen in the DFT calculations. Further discussion of the various optical transitions can be found in the Supplementary.

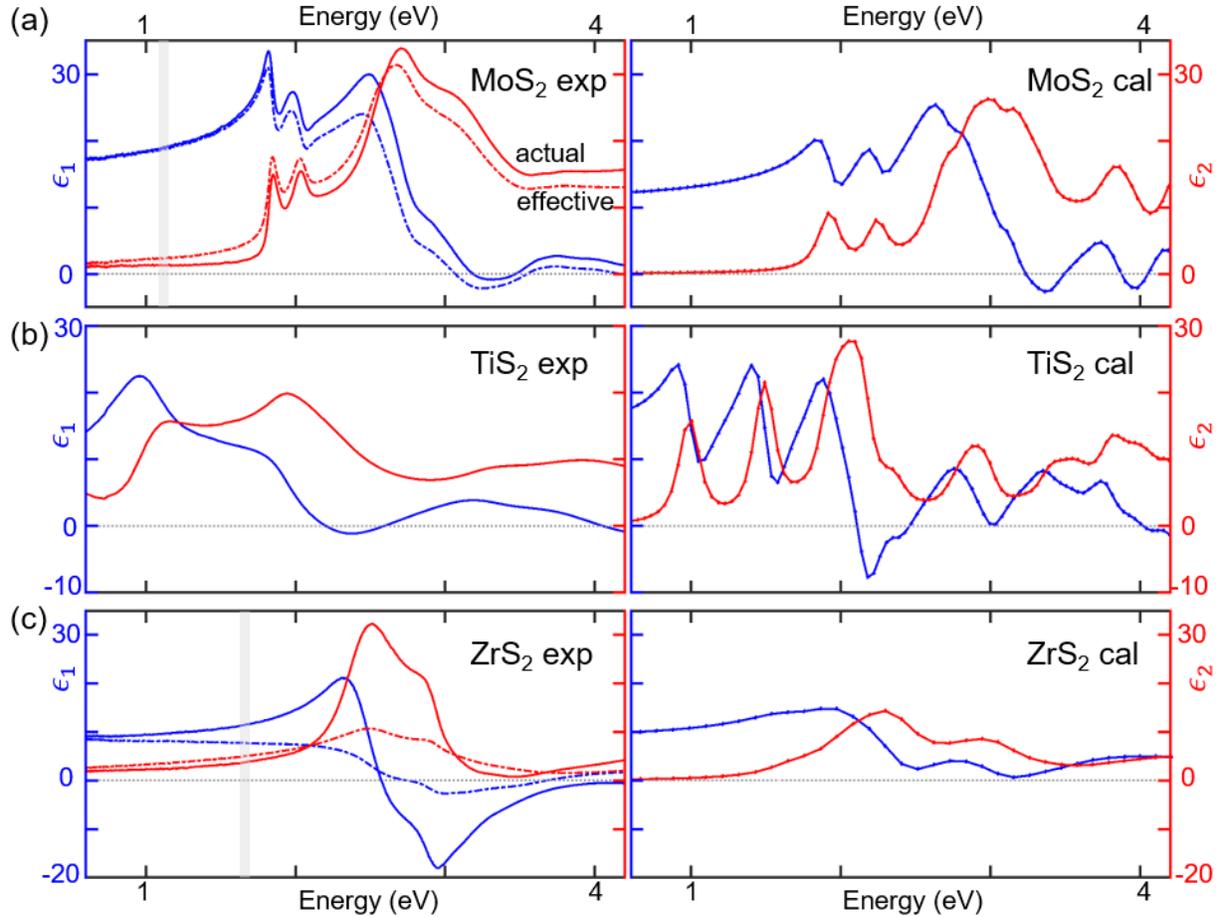

**Figure 2**: NIR-VIS complex relative permittivity ($\epsilon$) of sulfide TMDs. (Left column) Measured experimentally at room-temperature and AOI = 70° by spectroscopic ellipsometry. (Right column) Calculated by DFT. Results shown for (a) 2H-$MoS_2$, (b) 1T-$TiS_2$, and (c) 1T-$ZrS_2$. The indirect band gap of $MoS_2$ and $ZrS_2$ is indicated by light-gray lines. In the left column, the dashed lines for $MoS_2$ and $ZrS_2$ indicate the effective permittivity, not accounting for the native oxide; the solid lines show the actual permittivity, determined by analyzing the ellipsometry data taking into account the native oxide layers.

The presence of a native oxide layer affects the experimental results, particularly for regions where the optical loss of the TMD is expected small, such as below the band gap of $MoS_2$ and $ZrS_2$. We directly measured the thickness and composition of the native oxide using cross-sectional TEM (see Supplementary). On $MoS_2$ we find a rough surface, possibly including a native oxide, approximately 2 nm thick. On $ZrS_2$ we find a native oxide layer nearly 20 nm thick; similarly thick native oxide layers have been observed on $ZrSe_2$ [53]. On $TiS_2$ we saw no native oxide, within the imaging resolution of our experiment (~ 1 nm). Adding these overlayers to the optical model used to analyze the ellipsometry data significantly affects the extracted permittivity of $MoS_2$ and $ZrS_2$ (see Supplementary for modeling details). The solid lines in Figure 1 indicate the actual permittivity, determined by analyzing the ellipsometry data taking into account the native oxide layers.

The complex refractive index $(n - ik)$ is related to $\epsilon$ by

$$n = real(\sqrt{\epsilon}), \ k = -imaginary\left(\sqrt{\epsilon}\right). \qquad (4)$$

$k$ is related to the absorption coefficient ($\alpha = 4\pi k/\lambda$). In **Figure 3** we plot experimentally-measured $n$ and $k$ in the NIR spectral region 0.6 - 1.5 eV (827 – 2067 nm). All three materials have large $n$, comparable to or larger than that of silicon ($n_{NIR} \approx 3.4$), which is appealing for guiding NIR light in confined geometry. $MoS_2$ and $ZrS_2$ are indirect-band gap semiconductors and have low-loss in the NIR. $TiS_2$ is semi-metallic, and is predicted to have a small band-gap (~0.3-0.5 eV), and has higher loss.

The loss coefficients (*k*) determined by experiment and reported in **Figure 3** are conditional on the particular samples measured and on our optical modeling, and should be considered upper-bounds for these materials. The presence of a native oxide strongly influences the determined value of *k*. In **Figure 3** we indicate *k* determined from the effective permittivity (Equation 1 and 4), assuming no native oxide (dashed lines), and the value determined by optical modeling including native oxide thickness determined by TEM (solid lines). Taking the native oxide into account results in a lower value of the determined loss. Another important variable is the presence of defects, which contributes to below-band gap absorption and optical loss. Our samples are bulk crystals (including a naturally-occurring specimen of $MoS_2$) and definitely contain defects, including sulfur vacancies that contribute to NIR absorption [54]. Measurements on synthetic $MoS_2$ monolayers have shown lower loss in the NIR [40,41]. Our theoretical calculations are performed using models of perfect, defect-free crystals and predict substantially lower loss than the experiments (see **Figure 2**). As the science of processing TMD materials improves we will gain greater control over defects and can expect to have very low-loss TMDs for NIR applications.

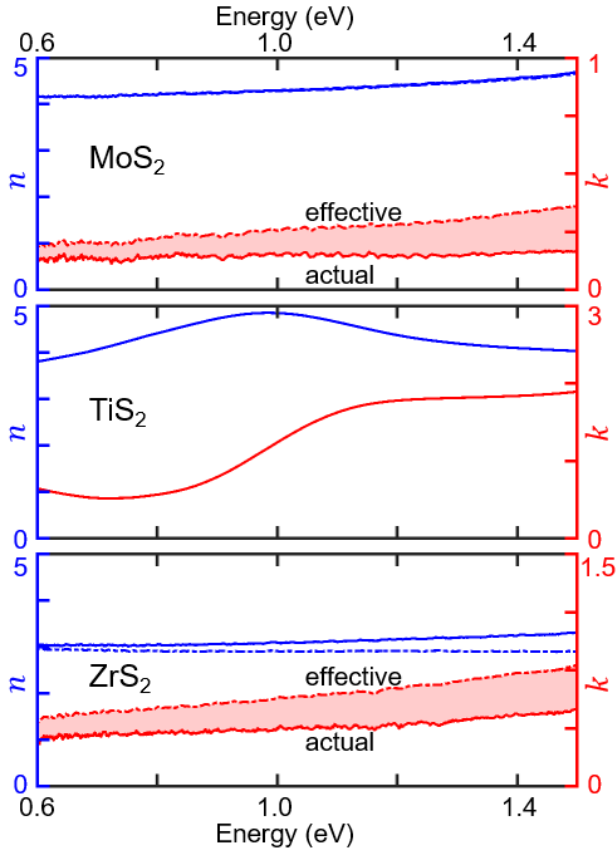

**Figure 3:** Experimentally-measured real ($n$) and imaginary ($k$) refractive index of MoS$_2$, TiS$_2$, and ZrS$_2$ in the NIR spectral region. The lower bound (solid lines) for MoS$_2$ and ZrS$_2$ are determined by modeling the ellipsometry data including native oxide layers. The upper bound (dashed lines) represent the effective permittivity, which ignores the native oxide. The red, shaded area represents experimental uncertainty due to potential mis-estimation of the native oxide thickness.

The results presented above are for in-plane polarized electric field. These measurements assume that the materials under study are birefringent, with the optical axis out-of-plane. We turn to MM measurements to explicitly measure the position of optical axis. MM provides the complete dielectric polarization of a material, and can measure cross-polarization components and depolarization effects due to surface layers [34,55]. The expected MM for a birefringent material, with the optical axis parallel to plane of incidence, is shown in **Figure 4** (top left). Due to symmetry, certain elements are expected to be zero, while others should be repeated [33,35]. We carry out measurements on TiS$_2$ at AOI = 70°. With the experimental configuration of a rotating compensator and auto-retarder (using Woollam UV-NIR Vase), the last row of the MM is not measurable [31]. We find that the measurements match the expected symmetry of a MM corresponding to a birefringent material with optical axis in plane of incidence. Thus, the optical axis for TiS$_2$ points out-of-plane, with high in-plane rotational symmetry. MoS$_2$ also show similar behavior (measured through Semilab SE-2000, not shown here). We also convert the MM to Jones matrix (Equation (2) and see Supplementary) to explicitly measure the cross-polarization components, which can signify anisotropy and surface structure. The cross-polarization and depolarization components are small, consistent with minimal surface roughness.

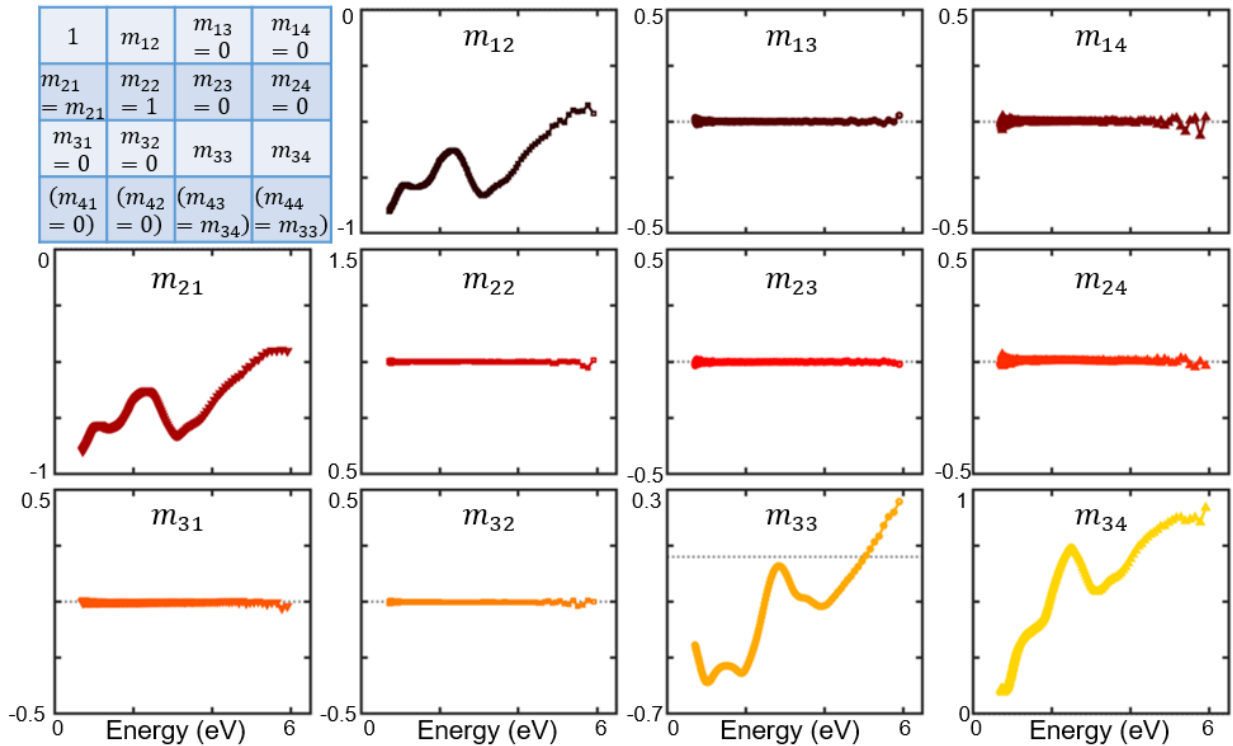

**Figure 4:** Mueller matrix (MM) ellipsometry on 1T-TiS$_2$. The table (top left) shows the symmetry expected from for a birefringent material with optical axis parallel to plane of incidence. The measurements ($m_{12}$ etc.) support this symmetry. The last row of MM is not measurable with the rotating compensator configuration used here. All components are normalized to $m_{11}$.

## 4. Results: Computed binary phase diagrams and phase-change properties

We focus on the NIR optical properties of sulfide TMDs because they offer lower optical loss (in NIR) than their selenide and telluride cousins, which have smaller band gap. Unfortunately, the energetic cost of switching between phases is also highest for pure sulfides (relative to selenides and tellurides) [25,27]. Alloying sulfide TMDs with different reference states could enable low-power switching. The thermodynamics of TMD alloys are not well-established, and no phase diagrams have been published for the $MoS_2$-$TiS_2$-$ZrS_2$ ternary system, or the subsidiary binary systems. Here we use DFT calculations to evaluate the likelihood of making binary alloys near the 2H-1T phase boundary.

We calculate the Gibbs free energy of pure phases and alloys using DFT and the quasi-harmonic approximation (QHA) method, aided by Phonopy code [56–58]. We use the QHA to calculate phonon spectra, which can be approximated as quantum harmonic oscillators at fixed lattice constant, and the vibrational entropy. To this entropy we add a configurational entropy term, although this term is dwarfed by the vibrational entropy. The calculation consists of 4 sequential steps: (1) calculate ground-state energy with fully-relaxed volume, (2) calculate energy as a function of volume for values around the ground state volume, (3) calculate phonon spectra at each volume, (4) calculate Gibbs free energy (see Supplementary). These four steps are akin to performing a Legendre transform from constant-volume and entropy, to constant-pressure and temperature. To model alloys, we use 2×2×2 (1T phase) and 2×2×1 (2H phase) supercells, thus each supercell contains 8 formula units. For each binary system we calculate five different compositions (*e.g.* $TiS_2$, $Ti_{0.75}Mo_{0.25}S_2$, $Ti_{0.50}Mo_{0.50}S_2$, $Ti_{0.25}Mo_{0.75}S_2$, $MoS_2$) in each of two phases (2H and 1T), with a suitable mesh as determined by a convergence test [59]. For a given sample, we calculate the dynamical matrix of each supercell using the Parlinski-Li-Kawazoe method, which combines the DFT results of a set of displaced supercells in which one ion is displaced by 0.01 Å. We derive the phonon spectrum by finding the non-zero eigenvalues of the resulting dynamical matrices. We then calculate the Helmholtz free energy (appropriate for a system at fixed temperature and volume) from the partition function of phonon vibrations. Finally, we calculate the Gibbs free energy (appropriate for a system at fixed temperature and pressure) by minimizing of the sum of Helmholtz free energy and *PV* (pressure times volume).

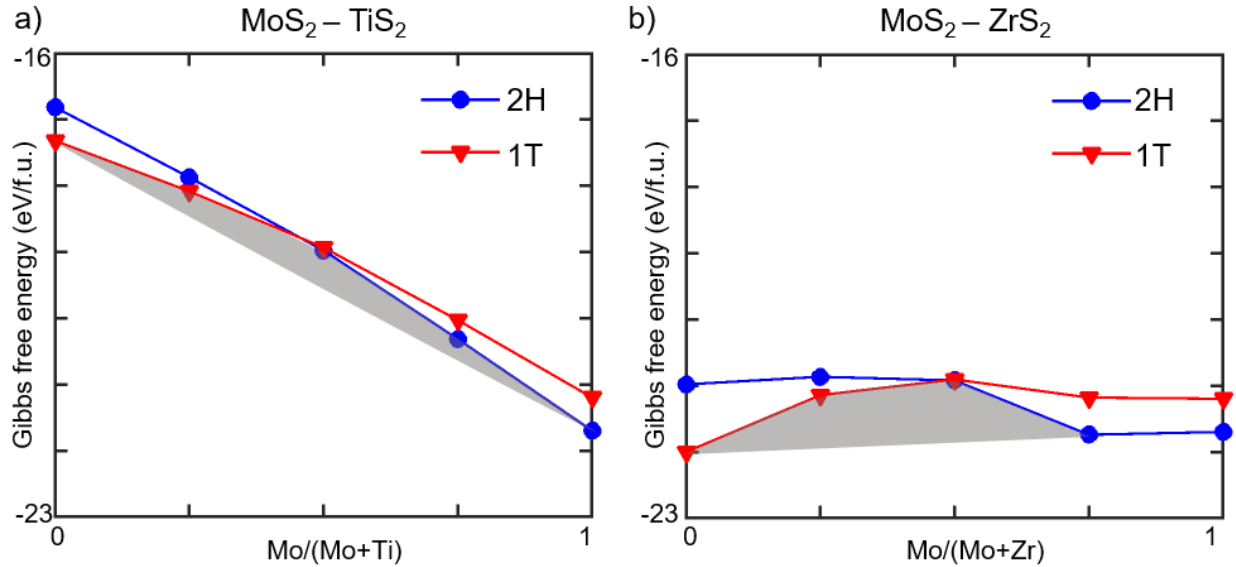

**Figure 5**: Theoretically-predicted Gibbs free energy-composition plots for the (a) $MoS_2$-$TiS_2$ and (b) $MoS_2$-$ZrS_2$ systems at 300 K. In both cases the alloys are thermodynamically unstable relative to decomposition into pure phases. However, the free energy above the convex hull (grey zone) is small, which suggests that kinetic stabilization will be possible, *e.g.* through low-temperature processing.

In **Figure 5** we show the calculated Gibbs free energy-composition curves for the $MoS_2$-$TiS_2$ and $MoS_2$-$ZrS_2$ systems at 300 K. The results are very similar at 1000 K (not shown here), although with an overall downward shift of ~ 1 eV/f.u (functional unit) relative to the data at 300 K. For both systems, the free energy curves for the 2H and 1T phases cross at an intermediate composition, which is suggestive of a phase boundary. For the $MoS_2$-$TiS_2$ system the curves are concave-downward and lie above the convex hull, which for this system is a straight line connecting the pure phases. Therefore, $MoS_2$-$TiS_2$ alloys will have a tendency to phase-separate at equilibrium. For the $MoS_2$-$ZrS_2$ we predict a solid solution in the 2H structure for $Mo_xZr_{1-x}S_2$, $x > 0.75$, and phase separation for more Zr-rich compositions. For both systems, the relatively small energy difference between the alloy curves and the convex hull (< 1 eV/f.u.) suggests that alloys may be kinetically-stabilized near the 2H-1T phase boundary. Future work should consider metastability and the kinetics of phase separation in TMD alloys systems, and a particular focus on low-temperature processing of TMD alloy thin films.

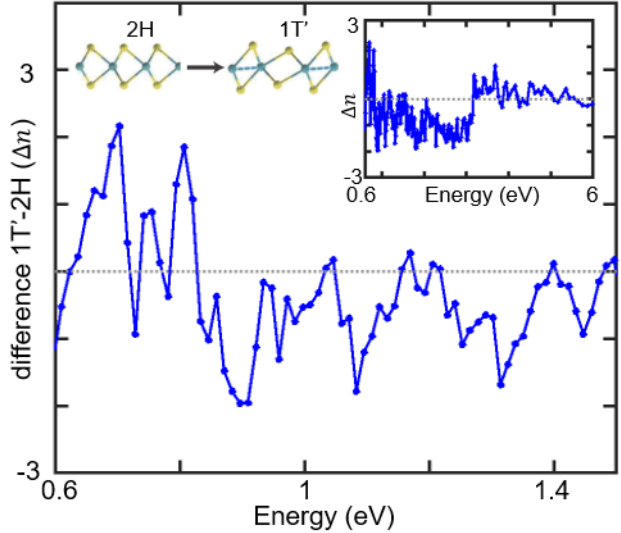

**Figure 6**: Theoretically-predicted refractive index difference ($\Delta n$) between the 1T' and 2H phases of monolayer $MoS_2$ in the NIR. (Inset, right) $\Delta n$ over a wider energy range. (Inset, left) Illustrated change of atomic structure for 2H-1T' phase transition.

We show (**Figures 2-3**) that sulfide TMDs are potentially useful materials for NIR photonics, with large refractive index and low-loss in the semiconducting phases, and that alloying (**Figure 5**) may be used to approach the 2H-1T phase boundary. We now address the question of the usefulness of sulfide TMDs as active materials for use in integrated photonics. In **Figure 6** we show the calculated refractive index difference ($\Delta n$) between the 2H and 1T' phases of monolayer $MoS_2$. We here show results for the 1T' phase instead of 1T because, according to our DFT calculations, 1T' has lower energy than 1T for monolayer $MoS_2$, and therefore 1T may spontaneously relax to 1T'. We find that $\Delta n$ is quite large, comparable to or larger than that realized by phase-change materials in the GST system. The theoretically-predicted spectral features are rather sharp, due to several factors: the calculations are performed for zero temperature, using a finite set of points in $k$-space, and do not include phonon-assisted transitions. In experimental reality these would likely be smoothed out, and we can expect $\Delta n \sim 1$ throughout the NIR.

*5. Conclusion*

We measure the complex optical constants of select sulfide TMDs in a spectral range from the visible to the NIR. The samples are single-crystals of 2H-$MoS_2$, 1T-$ZrS_2$, and 1T-$TiS_2$ and are chosen to represent prototypes of the 2H and 1T structure types. All materials have high index of refraction ($n \sim 3\text{-}4$), and MoS$_2$ and ZrS$_2$ feature low-loss in the NIR. We use Mueller matrix (MM) measurements to confirm that these materials are birefringent, as expected from their crystal structure and predicted by theory. The large refractive index and strong contrast in optical properties between the different structure types suggests a role for TMDs as phase-change materials for integrated photonics. Achieving this goal will require making materials that are thermodynamically-adjacent to a phase boundary. For sulfides this will likely require alloying. We use DFT to calculate free energy-composition curves for alloys of MoS$_2$, ZrS$_2$, and TiS$_2$. The alloy structural phases become energetically degenerate at intermediate compositions, at which martensitic switching may be possible, if the alloys are found to be metastable or kinetically-stable. Our calculations predict that transition metal sulfide phase-change materials will feature a large and useful $\Delta n \sim 1$ in the NIR, which is promising for controlling light in integrated photonic systems.

## Acknowledgments

This work was supported by an Office of Naval Research MURI through grant #N00014-17-1-2661. We acknowledge the use of facilities and instrumentation supported by NSF through the Massachusetts Institute of Technology Materials Research Science and Engineering Center DMR - 1419807. This work was performed in part at the Center for Nanoscale Systems (CNS), a member of the National Nanotechnology Coordinated Infrastructure Network (NNCI), which is supported by the National Science Foundation under NSF award no. 1541959. CNS is part of Harvard University. This work was performed in part at the Tufts Epitaxial Core Facility at Tufts University. We acknowledge assistance from the Department of Mineral Sciences, Smithsonian Institution. We acknowledge helpful discussions with Junho Choi, Yanwen Wu, Kevin Grossklaus, and John Byrnes.

**Supplementary information for**

**Near-infrared optical properties and predicted phase-change functionality of transition metal dichalcogenides**

Akshay Singh,[1] Yifei Li,[1] Balint Fodor,[2] Laszlo Makai,[2] Jian Zhou,[3] Haowei Xu,[3] Austin Akey,[4] Ju Li,[3] R. Jaramillo[1*]

1. Department of Materials Science and Engineering, Massachusetts Institute of Technology, Cambridge, MA, 02139, USA.

2. Semilab Semiconductor Physics Laboratory Co. Ltd., Budapest, Hungary

3. Department of Nuclear Science and Engineering, Massachusetts Institute of Technology, Cambridge, MA, 02139, USA.

4. Center for Nanoscale Systems, Harvard University, Cambridge, Massachusetts 02138, USA

* rjaramil@mit.edu

**I) Experimental setup**

Ellipsometry measurements were performed with two different instruments, Semilab SE-2000 and Woollam UV-NIR Vase. SE-2000 ellipsometry setup with a rotating compensator configuration is shown in **Figure S1(a)**. The SE-2000 system consist of the following elements in consecutive order: broadband white light source, polarizer, rotating compensator, micro-spot objective projecting the light onto the sample, analyzer, and finally the detection is performed by a CCD based multi-channel detector, or by a InGaAs photodiode array based detector for the NIR range. Woollam Vase is a ellipsometer with a rotating analyzer geometry, and an auto-retarder which functions analogous to a compensator. The focusing optics have a spot size of ~ 300 $\mu$m at 70° angle of incidence ($AOI$).

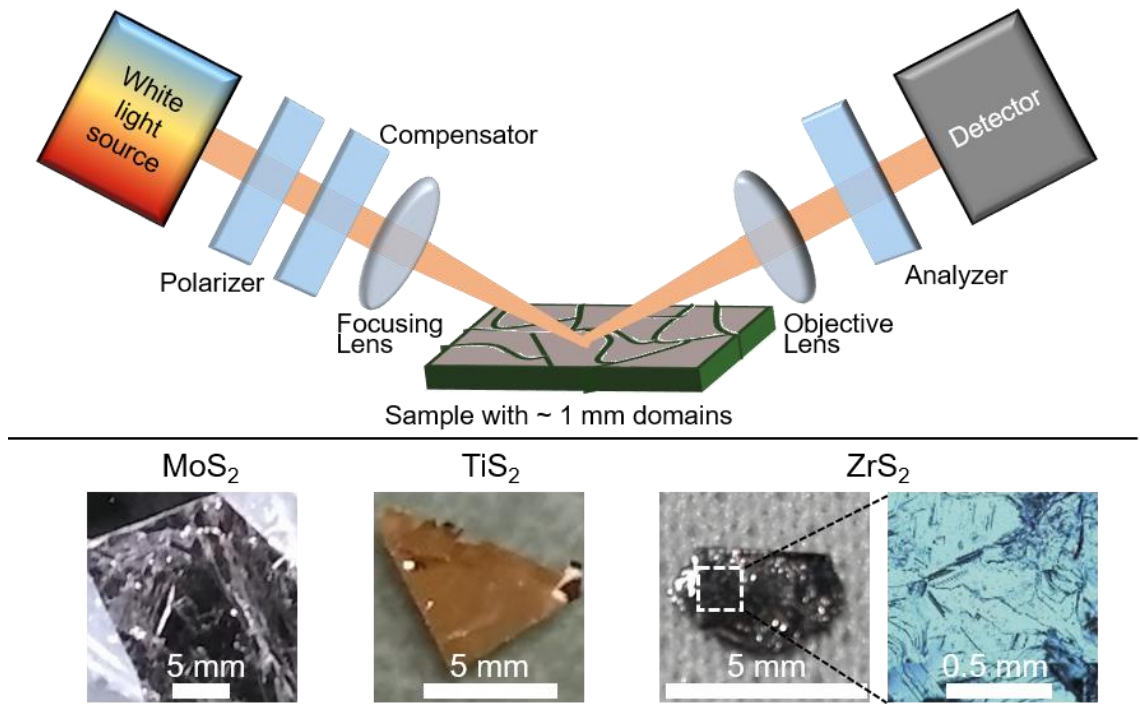

**Figure S1: (a)** Schematic of SE-2000 ellipsometry setup. Woollam VASE instrument performs similarly, but has an auto-retarder in place of the compensator, coupled with a rotating analyzer. **(b)** Low-magnification optical images of $MoS_2$, $TiS_2$ and $ZrS_2$ bulk crystals with different sized reflective facets. Higher magnification image (of dotted white box) of $ZrS_2$ is shown, clearly showing facets and heterogeneous nature of surface.

Optical images of bulk TMD crystals are shown in **Figure S1(b)**. Sufficiently thick pieces of naturally occurring $MoS_2$ (from Smithsonian Institution and 2D Semiconductors), were chosen to reduce back reflections. The surface had ~ 1 mm scale reflective facets, which necessitated the use of focusing optics. $TiS_2$ was purchased from 2D Semiconductors (grown using flux technique) and had large reflective domains ~ 5 mm. $ZrS_2$ was purchased from 2Dsemiconductors, grown using flux zone method, and had a surface similar to $MoS_2$, with reflective domains ~ 1 mm. $MoS_2$ was measured using both Semilab and Woollam ellipsometers, with similar results; $TiS_2$ was measured using Woollam, and $ZrS_2$ was measured using Semilab ellipsometer.

## II) Attributing peaks to different transitions

**Figure S2** shows the experimentally measured imaginary part of relative permittivity ($\epsilon_2$) for all TMDs under study (effective permittivity). The solid lines show the actual permittivity, determined by analyzing the ellipsometry data taking into account the native oxide layers, using the model explained in Section IV. Observable optical transitions are labelled by transition metal and subscript (for example A-exciton transition in MoS$_2$ ~ $M_A$). Then in Table S1, we attribute spectral peaks in $\epsilon_2$ to different optical transitions observed in earlier literature.

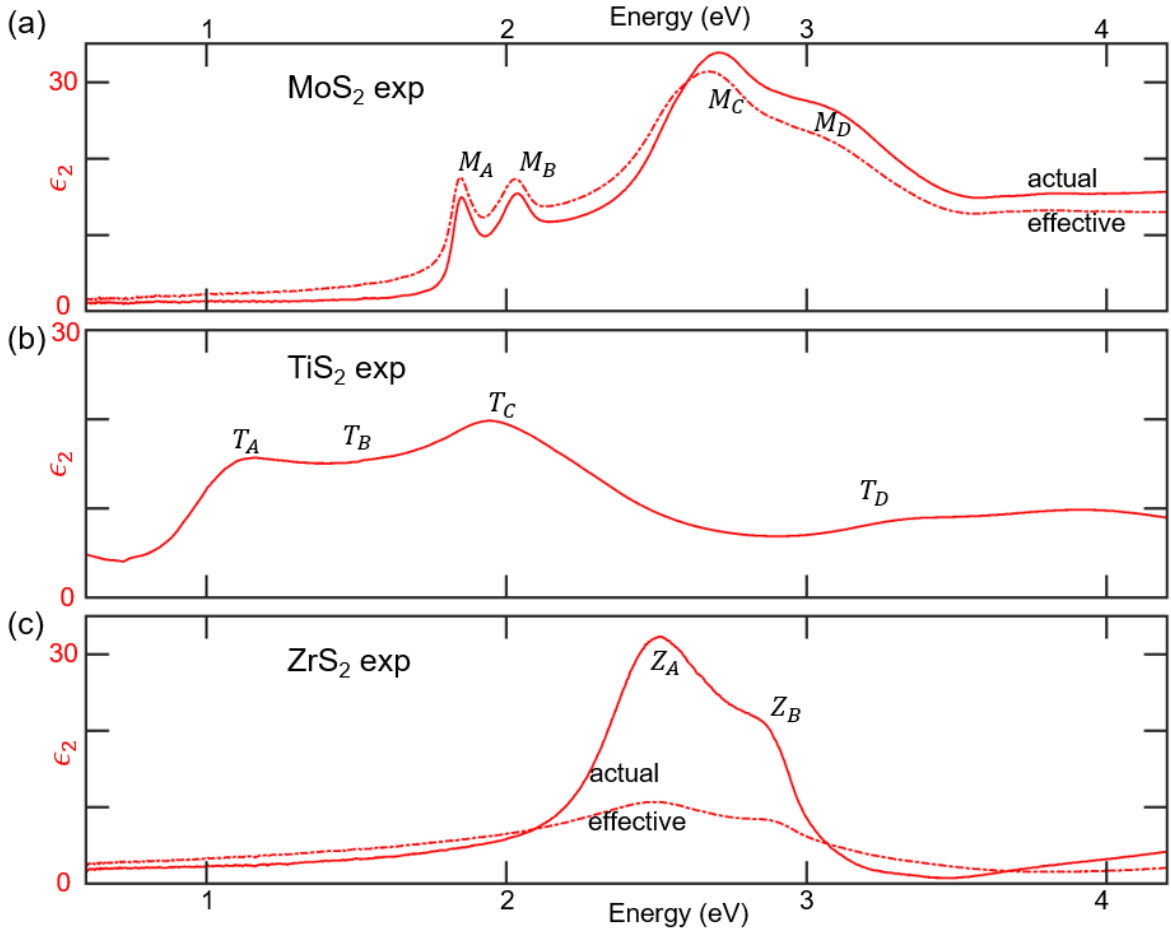

**Figure S2:** Imaginary part of experimentally measured relative permittivity for **(a)** MoS$_2$, **(b)** TiS$_2$ and **(c)** ZrS$_2$ ($AOI = 70°$). The different optical transitions are indicated by the transition metal and subscript. The solid lines show the actual permittivity, determined by analyzing the ellipsometry data taking into account the native oxide layers; the dashed lines show the effective permittivity, not accounting for the native oxide.

| TMD | Peak and energy (eV) | Possible origin or literature |
|---|---|---|
| MoS$_2$ | A ~ 1.8 eV | A Exciton, Ref [1] |
| | B ~ 2.1 eV | B Exciton, Ref [1] |

|  | C ~ 2.7 eV | C Exciton (band-nesting), Ref [1] |
|---|---|---|
|  | D ~ 3.1 eV | D Exciton, Ref [1] |
| TiS$_2$ | A ~ 1.1 eV | Ref [2] |
|  | B ~ 1.5 eV | Ref [2] |
|  | C ~ 1.9 eV | Possible 2 peaks, Ref [2] |
|  | D ~ 3.2 eV | Possible 2 peaks, Ref [2] |
| ZrS$_2$ | A ~ 2.5 eV | Possible 2 excitonic peaks, Ref [2] |
|  | B ~ 2.9 eV | Possible 2 excitonic peaks, Ref [2] |

**Table S1:** Spectral peaks observed in experimentally measured $\epsilon_2$ attributed to different optical transitions for MoS$_2$, TiS$_2$ and ZrS$_2$. The experimental observations are also compared to previous literature.

### III) Transmission Electron Microscopy (TEM) characterization of surface

High resolution TEM (HRTEM) and Scanning TEM (STEM) are used to characterize the surface of bulk crystals. For TEM, cross-sectional samples are prepared via a gallium focused ion beam (FIB). A thin (~ 100 nm) amorphous carbon layer is deposited on top of the crystal for protection during subsequent FIB steps. Dark field (DF) STEM performed on MoS$_2$ (**Figure S3a**) indicates a slightly rough interface between carbon and underlying layered structure. HRTEM performed on TiS$_2$ (**Figure S3b**) shows a very sharp interface between FIB-deposited carbon and underlying crystals, and no overlayers. The ZrS$_2$ sample on the contrary (**Figure S3c**), shows a thick amorphous layer on top of the pristine crystalline structure.

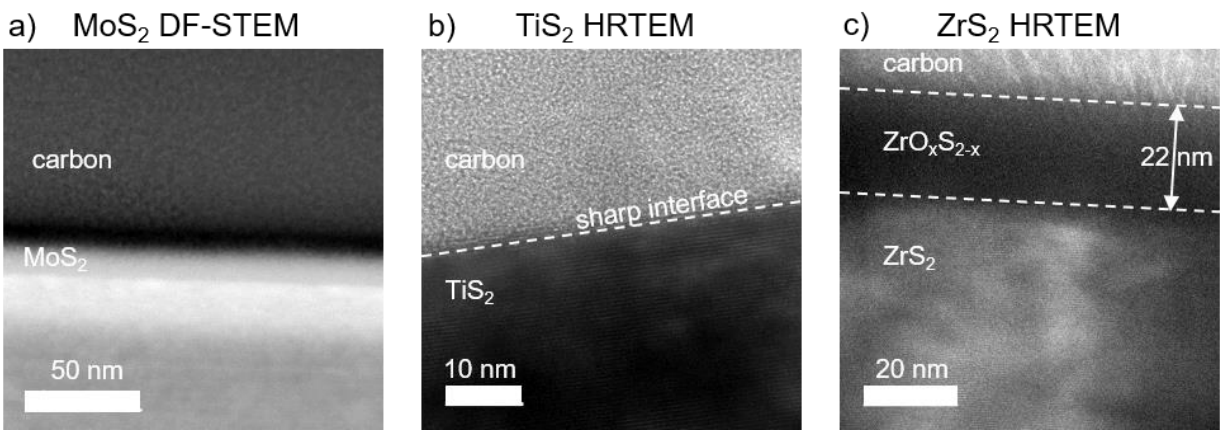

**Figure S3**: High-resolution TEM and STEM for sulfides. The carbon is deposited by focused ion beam instrument for protection of underlying crystalline material. a) Dark-field (DF) STEM for $MoS_2$ sample shows slightly rough interface between FIB-deposited carbon and underlying $MoS_2$. b) HRTEM for $TiS_2$ sample shows sharp interface between deposited carbon and underlying crystalline $TiS_2$. c) HRTEM for $ZrS_2$ sample shows amorphous overlayer on underlying crystalline $ZrS_2$.

**IIIb) Energy Dispersive X-ray Spectroscopy (EDS) of MoS₂ and ZrS₂ top surfaces**

To characterize the elemental composition of observed amorphous layers and top surface for $MoS_2$ and $ZrS_2$, we performed high resolution Energy Dispersive X-ray Spectroscopy (EDS) on the cross-sectional samples. In **Figure S4**, integrated intensities of characteristic X-ray peaks corresponding to carbon, transition metal (molybdenum/zirconium), sulfur and oxygen are plotted. For $MoS_2$, the interface seems fairly sharp (**Figure S4a**), and there is a distinct lack of oxygen in the underlying crystals. For $ZrS_2$, the amorphous layer measured via HRTEM, has a higher (lower) value of oxygen (sulfur/zirconium) compared to underlying pristine layers (**Figure S4b**).

Qualitatively, we assign an average composition of 50% sulfur – 50% oxygen to this layer.

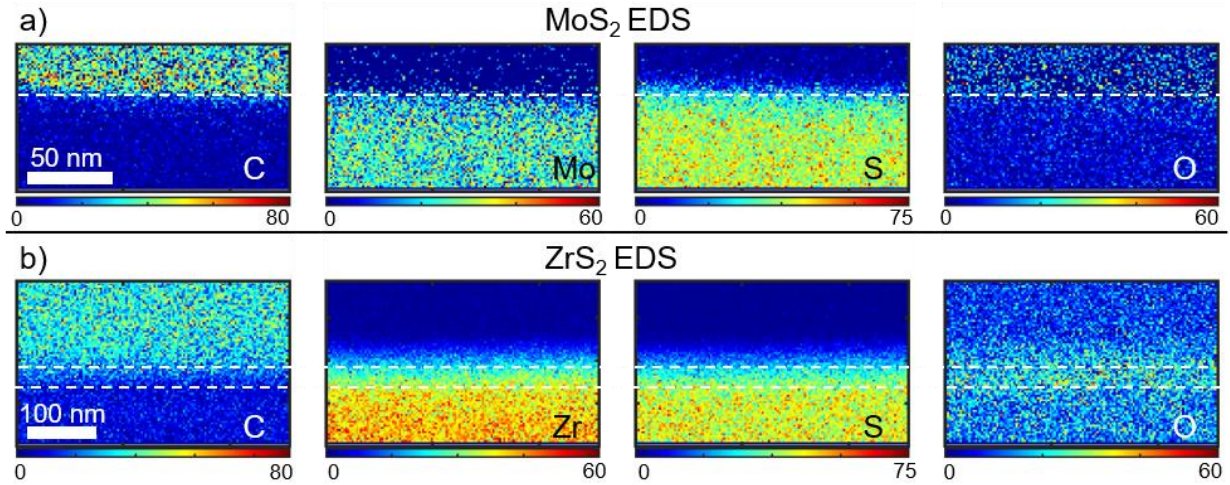

**Figure S4:** Energy Dispersive X-ray Spectroscopy (EDS) characterization of cross-sectional samples of a) MoS$_2$ and b) ZrS$_2$. The $K_{\alpha 1}$ lines are characteristic of the different elements (carbon, molybdenum/zirconium, sulfur and oxygen). The X-ray peaks are integrated to improve signal to noise.

## IV) Oxide and roughness modeling

TEM measurements show the presence of a moderately rough layer for MoS$_2$ (confirmed by atomic force microscopy, not shown here). The roughness is quantified as ~ 2 nm. Thus equation 3 of main text can-not be simply used, and an optical model incorporating different layers needs to be defined. We define an optical model for the MoS$_2$ sample in **Figure S5a**. Similarly, an optical model for the ZrS$_2$ sample, incorporating a 20 nm overlayer is shown in **Figure S5b**. The modeling is performed using Semilab and Woollam software, and give similar results. Results of the modeling are shown in **Figure S2**, and **Figure 2** and **3** of main text.

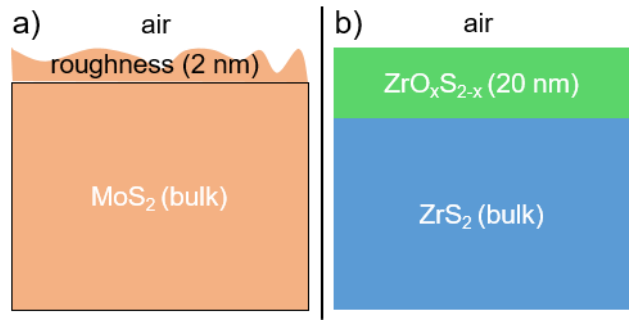

**Figure S5:** Schematic of the layer structure used for modeling raw ellipsometry data. a) $MoS_2$ data modeling incorporates a roughness layer of thickness ~ 2 nm on top of bulk $MoS_2$. b) $ZrS_2$ data modeling incorporates a zirconium oxy-sulfide layer (thickness ~ 20 nm) on top of bulk $ZrS_2$.

### V) $TiS_2$ with and without focusing optics

The focusing optics are necessary for a small spot size to avoid covering multiple small domains (with different tilts), which can cause mixing of polarizations. However, focusing optics can result in small amount of depolarization, and a change in the measured relative permittivity [3]. In **Figure S6**, we show that the measurements on $TiS_2$ using broad beam mode (3 mm spot size) and focusing beam mode (0.3 mm spot size) are quantitatively similar. Thus the focusing mode can be used, without introducing error in the measurements. Note that such a comparison is only possible for $TiS_2$, since $MoS_2$ and $ZrS_2$ have very small (~ 1 mm) reflective domains.

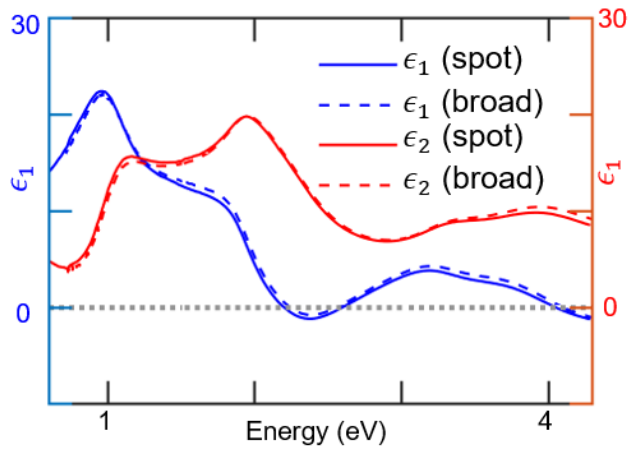

**Figure S6:** Comparison of relative permittivity measured via focused beam (spot) and broad beam (broad) for $TiS_2$ ($AOI = 70°$). The measurements are quantitatively similar, thus demonstrating focusing mode is suitable for ellipsometric measurements.

### VI) Mueller matrix (MM) to equivalent Jones matrix (S)

MM provides the complete polarization state of a material. In the case of no depolarization, a MM can be converted into an equivalent Jones matrix ($S$) which explicitly indicates any cross-polarization conversion, and is simpler to understand

$$MM \rightarrow S = \begin{bmatrix} r_{pp} & r_{ps} \\ r_{sp} & r_{ss} \end{bmatrix} \Rightarrow \rho = \begin{bmatrix} \rho_{pp} & \rho_{ps} \\ \rho_{sp} & 1 \end{bmatrix} \quad (1)$$

, where the different components were described in equation (2) of main text.

This conversion $MM \rightarrow S$ can be quantified with a quality factor, related to depolarization. We find that quality factor (depolarization) is indeed ~ 0 for TiS$_2$ (**Figure S7(a)**), indicating that the measured MM is a pure MM [4,5]. Thus the conversion $MM \rightarrow S$ can be carried out (**Figure S7(b)**). The cross-polarized (off-diagonal) components of $S$ are ~ 0, indicating minimal polarization conversion.

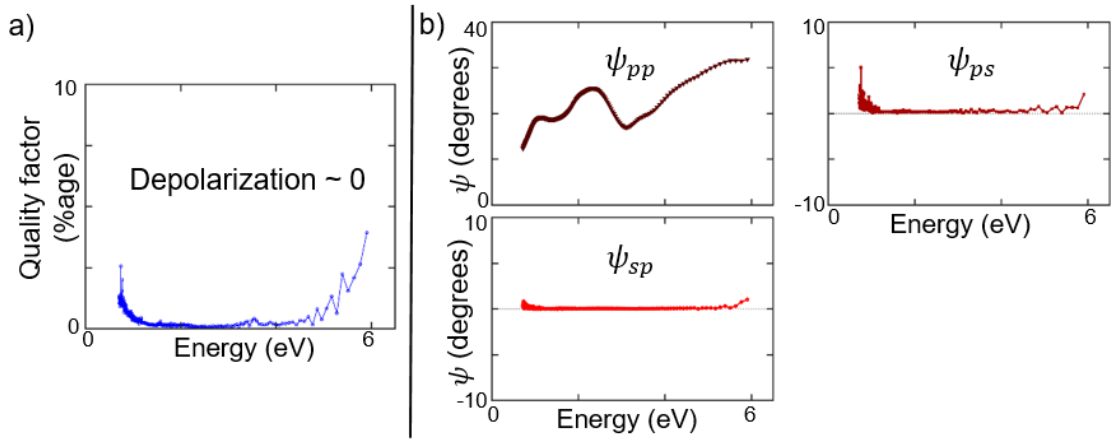

**Figure S7:** Mueller matrix ($MM$) to equivalent Jones matrix conversion ($S$) for TiS$_2$ ($AOI = 70°$). (a) Quality factor quantifying the $MM \rightarrow S$ conversion. The quality factor is analogous to a depolarization measurement. (b) Diagonal and off-diagonal components of equivalent $S$. Here, only $\psi$ is plotted. All components are normalized to $\rho_{ss}$. Off-diagonal components are ~ 0.

## VII) Theoretical calculations (DFT) for out-of-plane relative permittivity

Layered materials are expected to have large anisotropy due to the weak inter-layer bonding. Using DFT, we calculate the out-of-plane relative permittivity ($\epsilon_\parallel$) for all three TMDs under study. In **Figure S8**, $\epsilon_\parallel$ and $\epsilon_\perp$ are plotted together for comparison. The calculated $\epsilon_\parallel$ is much smaller than $\epsilon_\perp$ (compare with **Figure S2**, and Main Text **Figure 2**). Thus, we predict a large anisotropy between in-plane and out-of-plane directions, and these layered materials can be explored for optical modulators and polarization converters.

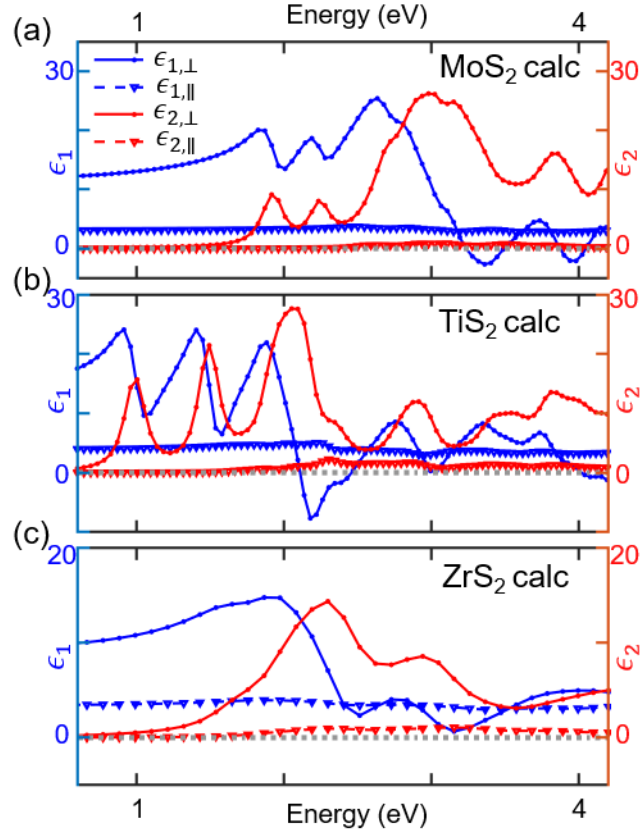

**Figure S8:** Comparison of calculated relative permittivity in ($\epsilon_\perp$, solid line with circular markers) and out-of-plane ($\epsilon_\parallel$, dotted line with triangular markers) for **(a)** $MoS_2$, **(b)** $TiS_2$ and **(c)** $ZrS_2$. The $\epsilon_\parallel$ is much smaller than $\epsilon_\perp$, and gives rise to large anisotropy in refractive index.

### VIII) Multiple angle of incidence (MAI) measurements

To separate $\epsilon_\perp$ and $\epsilon_\parallel$, measurements at more than one $AOI$ are needed. Measurements at low $AOI$ are insensitive to $\epsilon_\parallel$, and high $AOI$ is usually needed to measure $\epsilon_\parallel$. We carry out MAI measurements for $MoS_2$ and $TiS_2$, and display real part of effective (measured) relative permittivity ($\epsilon_1$) in **Figure S9**. DFT calculations suggest a low $\epsilon_\parallel$, and suggest an increase in effective $\epsilon$ with increase in angles. The lack of measured changes in effective $\epsilon_1$ for higher $AOI$ (for $MoS_2$, $TiS_2$) is thus puzzling. However, after explicitly measuring the direction of optical axis (via MM), we realize that even at high $AOI$ (for such high index materials), the measurement of effective $\epsilon$ is only weakly dependent on $\epsilon_\parallel$. Thus, MAI measurements are unable to uncouple $\epsilon_\perp$ and $\epsilon_\parallel$ due to high index and absorption. In the next section, we calculate effective $\epsilon$ for a range of $\epsilon_\perp$ and $\epsilon_\parallel$, and for different $AOI$, and predict changes of only a few percent with different $AOI$

(below signal to noise of our setup). A way forward is to perform MAI measurements on the side-plane of polished and thick TMD crystals, where the optical axis will depend on sample rotation, and anisotropy will be extractable [6]. Such measurements are however, beyond the scope of this paper.

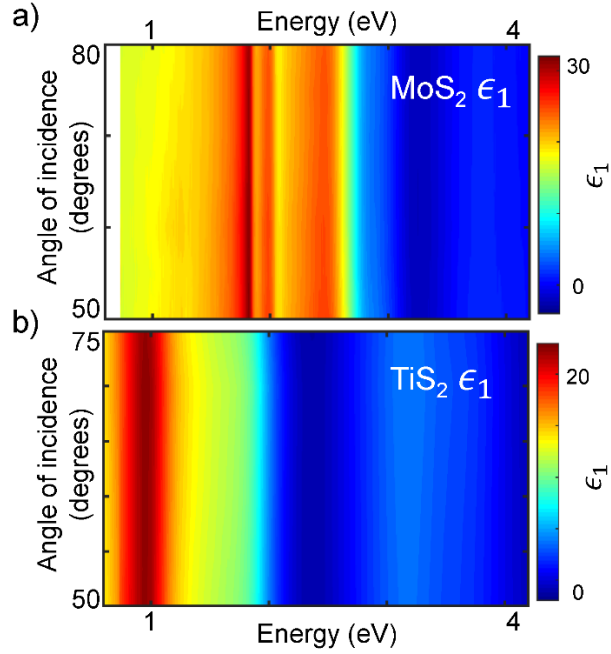

**Figure S9:** MAI measurements, plotting $\epsilon_1$ for (a) MoS$_2$ and (b) TiS$_2$. No noticeable change is measured between low and high incidence angles.

**IX) Effective $\epsilon$ for a range of $\epsilon_\perp$ and $\epsilon_\parallel$, and for different angle of incidence (AOI)**

MAI measurements did not reveal a significant change in effective $\epsilon$ for any TMD (MoS$_2$, TiS$_2$, ZrS$_2$). With the optical axis position confirmed to be out-of-plane, we carry out calculations for $n_{effective}(=\sqrt{\epsilon})$ for a range of in-plane ($n_\perp$) and out-of-plane ($n_\parallel$) refractive indices. Fresnel coefficients for a uniaxial material can be written [7] as

$$r_{pp} = \frac{n_\perp * n_\parallel * cos(\Phi) - n_i * \left(\left(n_\parallel^2 - (n_i * sin(\Phi))^2\right)^{\frac{1}{2}}\right)}{n_\perp * n_\parallel * cos(\Phi) + n_i * \left(\left(n_\parallel^2 - (sin(\Phi))^2\right)^{\frac{1}{2}}\right)} \qquad (2)$$

$$r_{ss} = \frac{n_i * cos(\Phi) - \left(n_\perp^2 - (n_i * sin(\Phi))^2\right)^{\frac{1}{2}}}{n_i * cos(\Phi) + \left(n_\perp^2 - (n_i * sin(\Phi))^2\right)^{\frac{1}{2}}} \quad (3)$$

, where $n_i, n_\perp, n_\parallel$ are ambient, in-plane and out-of-plane refractive (real part) indices respectively, and $\Phi$ is angle of incidence. Subsequently, an $n_{effective}$ can be defined and calculated using the ellipsometric ratio ($\rho = r_{pp}/r_{ss}$) followed by $\epsilon$ using

$$\epsilon = sin^2(\Phi) * \left(1 + tan^2(\Phi) * \left(\frac{1-\rho}{1+\rho}\right)^2\right) \Rightarrow n_{effective} = real(\sqrt{\epsilon}) \quad (4)$$

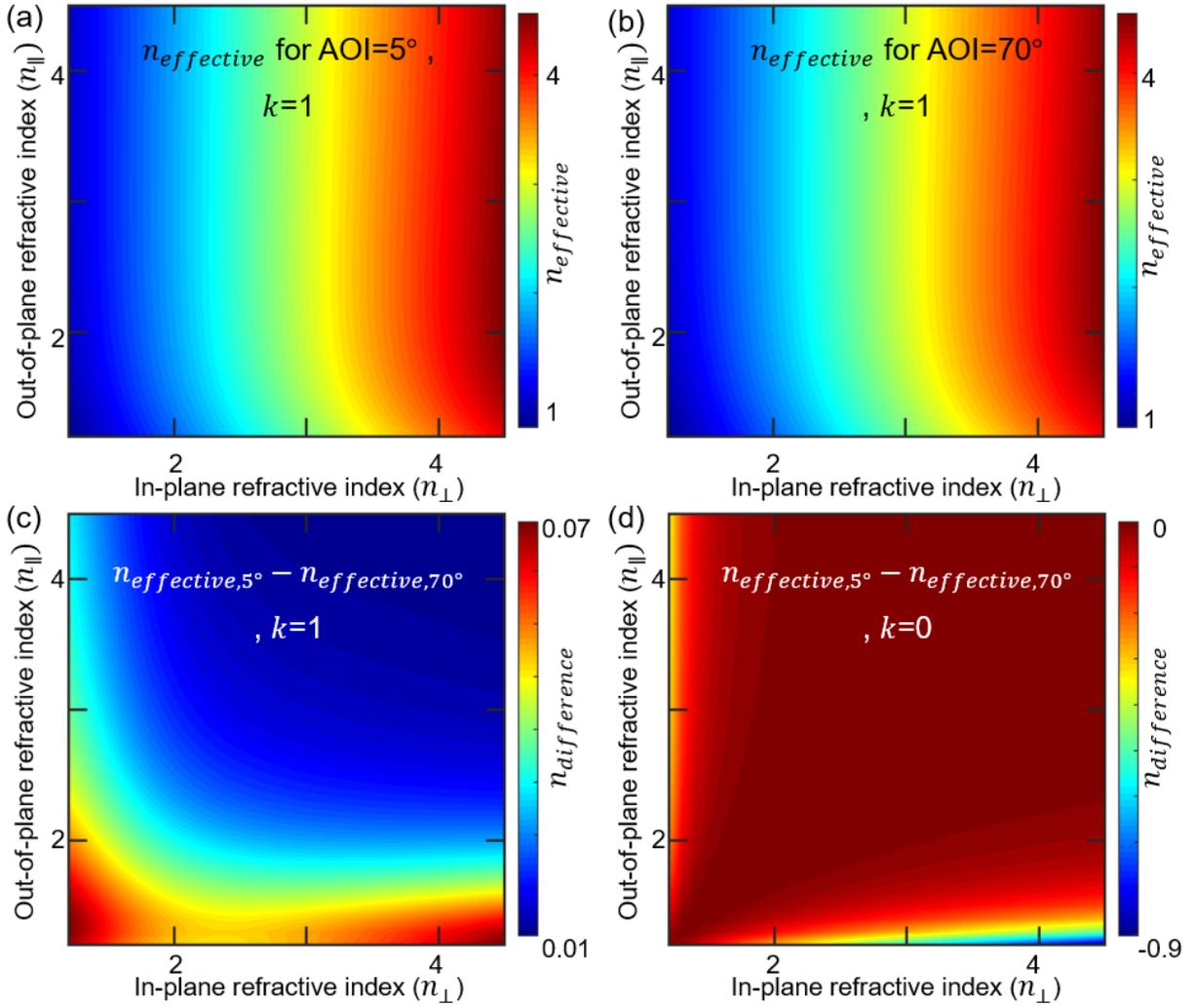

**Figure S10:** Calculation of $n_{effective}$ for a range of $n_\perp$ and $n_\parallel$, and for different AOI. **(a)** $k = 1$, $\Phi = 5°$. **(b)** $k = 1$, $\Phi = 70°$. **(c)** $n_{effective,5°} - n_{effective,70°}$, for $k = 1$. **(d)** $n_{effective,5°} - n_{effective,70°}$, for $k = 0$.

In **Figure S10**, $n_{effective}$ for a range of $n_\perp$ and $n_\parallel$ is plotted, using realistic values of imaginary part of refractive index ($k$). Two representative AOI of 5° and 70° (**Figure S10(a,b)**) are chosen to illustrate the change of $n_{effective}$ with different $n_\perp$ and $n_\parallel$, where $k = 1$ is fixed. However, the difference between $n_{effective,5°}$ and $n_{effective,70°}$ is very small (**Figure S10(c)**) and within experimental confidence limits. Thus the multiple AOI measurement may not be suitable for these high-index and absorptive materials. We also calculate the index difference for the case $k = 0$, and observe a measurable (~ 0.4) change in $n_{effective}$. Thus, for a transparent highly anisotropic material, multiple AOI measurements may be able to measure the anisotropy.

## X) Density function theory (DFT) calculations

Some details of DFT are provided in the main text, and are supplemented here. For the dielectric function calculation, 10×10×5 Monkhorst-Pack k point mesh and fully relaxed unit cell are used for reasonable accuracy [8]. Following our previous work, ion-clamped dielectric functions are predicted by Random Phase Approximation (RPA) [9,10] with the equation:

$$\epsilon_{\alpha\alpha}(\omega) = 1 - \lim_{q \to 0} \left(\frac{4\pi e^2}{q^2 \Omega}\right) \sum_{c,v,k} w_k \times \frac{|\langle u_{v,k}|u_{c,k+q_\alpha}\rangle|^2}{E_{c,k} - E_{v,k} - \hbar\omega - i\zeta} \quad (\alpha = x, y, z)$$

, where $c$ and $v$ indicate the conduction band states and valence band states respectively, $|u_{n,k}\rangle$ is the cell-periodic part of the wave functions of the band-$n$ at $k$, $\Omega$ is the volume of the simulation supercell, $w_k$ is the $k$-point weight and $\zeta$ is a phenomenological damping parameter (taken to be 0.025 eV).

The results are in qualitative agreement with the experiments results (**Figure 2** of main text). However, it should be noted the DFT calculated electronic band structure (Kohn-Sham eigenvalues) has systematic errors, and methods beyond the independent particle picture are required for better accuracy.

## XI) Phase diagram calculation

The steps involved in phase diagram calculations are provided below:

*1) Fully relaxed volume:* Alloy metal sulfides are simulated in 2×2×2 (1T phase) and 2×2×1 (2H phase) supercells, thus each supercell contains 8 formula units. By varying the composition of 8 metal atoms in each supercell, 5 stoichometries ($TiS_2$, $Ti_{0.75}Mo_{0.25}S_2$, $Ti_{0.50}Mo_{0.50}S_2$, $Ti_{0.25}Mo_{0.75}S_2$, $MoS_2$) in two phases (2H and 1T) are investigated. A Monkhorst-Pack k-point mesh of size 3×3×2 has been chosen, whose convergence has been tested up to 10×10×5. Each supercell is fully relaxed without any symmetry constraints by the conjugate gradient method, and the initial positions of ions are manually perturbed to promote the gradient descent.

*2) Relaxation with Fixed volume:* At given phase and stoichiometry, every fully relaxed supercell is strained by 0.5% in each vector, up to 2.5% in both compression and tension, to generate 11 different volume points. Each point is relaxed under similar condition but with fixed volume.

*3) Phonon spectra:* Under fixed phase, stoichiometry and volume, the dynamical matrix of each supercell is calculated by Parlinski-Li-Kawazoe method, combining the DFT simulation results of a set of displaced supercells in which one ion is displaced by 0.01 Å. The phonon spectra and density is derived by solving non-zero eigenvalue dynamical matrices. Then Helmholtz free energy ($F$) is the direct result of the partition function ($Z$) of phonon vibrations:

$$Z = \prod_{qv} \frac{\exp(-\hbar\omega(qv)/(2k_B T))}{1 - \exp(-\hbar\omega(qv)/(k_B T))}$$

$$F = -k_B T \ln Z = \frac{1}{2}\sum_{qv} \hbar\omega(qv) + k_B T \sum_{qv} \ln[1 - exp(-\hbar\omega(qv)/(k_B T))]$$

, where $q$ is the wave vector and $v$ is the band index. All the phonon spectra calculations use 15×15×15 $q$-point mesh.

*4) Calculate the Gibbs free energy:* Gibbs free energy (at given phase and stoichiometry under 0 Pa pressure) are calculated by minimization of the sum of Helmholtz free energy and $pV$, which only depends on volume under fixed temperature.

$$G(T,p) = \min_V [U(V) + F_{phonon}(T,V) + pV]$$

, where $U(V)$ is the DFT calculated ground state energy. The configuration entropy term from Mo-Ti alloy is also added to the final results, but the contribution is in the scale of 0.1 eV, which is negligible.

$$S = -k_B \sum_{n=1}^{W} P_n \ln P_n$$

Also, the contribution of $pV$ term is negligible since it is ~ 1 meV when pressure ~ 1 atmosphere.